# Influence of shear rate and surface chemistry on thrombus formation in micro-crevice

Mansur Zhussupbekov [a], Wei-Tao Wu [b], Megan A. Jamiolkowski [c], Mehrdad Massoudi [d], James F. Antaki [a]

[a] Meinig School of Biomedical Engineering, Cornell University, Ithaca, NY, USA
[b] School of Mechanical Engineering, Nanjing University of Science and Technology, Nanjing, China
[c] U.S. Food and Drug Administration (FDA), Center for Devices and Radiological Health (CDRH), Office of Science and Engineering Laboratories (OSEL), Silver Spring, Maryland, USA
[d] U.S. Department of Energy, National Energy Technology Laboratory (NETL), Pittsburgh, PA, USA

*Address for correspondence:*
James F. Antaki, PhD
Weill Hall 109
Meinig School of Biomedical Engineering
Cornell University
Ithaca, NY 14853
Phone: 607-255-0726
Email: antaki@cornell.edu

**Abstract**
Thromboembolic complications remain a central issue in management of patients on mechanical circulatory support. Despite the best practices employed in design and manufacturing of modern ventricular assist devices, complexity and modular nature of these systems often introduces internal steps and crevices in the flow path which can serve as nidus for thrombus formation. Thrombotic potential is influenced by multiple factors including the characteristics of the flow and surface chemistry of the biomaterial. This study explored these elements in the setting of blood flow over a micro-crevice using a multi-constituent numerical model of thrombosis. The simulations reproduced the platelet deposition patterns observed experimentally and elucidated the role of flow, shear rate, and surface chemistry in shaping the deposition. The results offer insights for design and operation of blood-contacting devices.

**1. Introduction**
Left ventricular assist devices (VAD) are a widely accepted and often indispensable treatment option for end-stage heart failure (Kirklin et al., 2015; Teuteberg et al., 2020). However, VAD therapy remains plagued by serious adverse events (Acharya et al., 2017; Kirklin et al., 2017; Li et al., 2020; Nguyen et al., 2016) with stroke emerging as the leading cause of death (Kormos et al., 2019; Li and Mahr, 2019; Mcilvennan et al., 2019). VAD-associated stroke often occurs in conjunction with pump thrombosis (Cho et al., 2019).

Complexity of blood pumps often mandates that these devices are assembled from multiple parts. This can introduce internal steps, crevices, and irregularities at the junction sites. In blood pumps with mechanical bearings such as Heartmate II (HMII, Abbott Laboratories-previously Thoratec Corp., Pleasanton, CA), the junction between the rotating and stationary parts creates a discontinuity in the blood-contacting surface. Thrombus growth is commonly observed at the inlet ball-and-cup bearing of the HMII as shown in Fig. 1(a) (Bhamidipati et al., 2010; Kittipibul et al., 2020; Kreuziger et al., 2018; Mokadam et al., 2011; Rowlands et al., 2020; Tsubota et al., 2017).

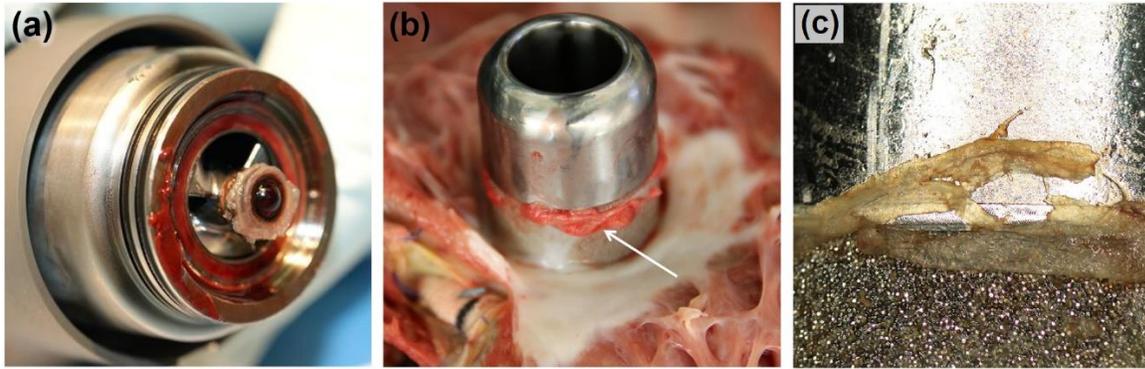

**Fig. 1.** (a) Thrombus found on the HMII inlet bearing. Image courtesy of Dr. Francis D. Pagani. (b) Thrombus at the junction between the smooth and sintered portions of the inflow cannula of the HVAD. Reprinted with permission from Elsevier (Strickland et al., 2016). (c) Close-up view of a thrombus at the smooth-sintered interface on the HVAD inflow cannula. Reprinted with permission from Elsevier (Glass et al., 2019).

Local variations in surface treatment and texture of blood-contacting surfaces can also lead to micro-scale irregularities and crevices at the sites of transition. In the HeartWare HVAD (Medtronic, Inc., Minneapolis MN), the exterior surface of the inflow cannula features a textured section of sintered titanium microspheres that covers approximately half of the length of the cannula. While this surface modification had succeeded in promoting tissue ingrowth as originally intended (Najjar et al., 2014), the transition from the sintered section to the smooth, polished section has later emerged as a critical site for thrombus development (Glass et al., 2019; Kaufmann et al., 2018; Pappalardo et al., 2020; Potapov et al., 2017; Strickland et al., 2016). Fig. 1(b) and (c) show thrombi confined to the ring of discontinuity at the smooth-sintered interface (Glass et al., 2019; Strickland et al., 2016).

Motivated by the above clinical observations, Jamiolkowski et al. (2016) conducted a microfluidic experimental study of human platelet deposition on titanium alloy (Ti6Al4V), a biomaterial commonly used in VADs (Sin et al., 2009; Yamane, 2016), in the presence of crevices approximately 50 to 150 μm in width. To observe the platelet deposition in real time, they used reconstituted fresh human blood with hemoglobin-depleted red blood cells (RBC ghosts) (Bozzo et al., 2001; Jamiolkowski et al., 2015; Schwoch and Passow, 1973) and fluorescently labeled platelets. Complimentary numerical simulations were published by Wu et al. (2017) employing a multi-constituent model of thrombosis. These simulations replicated experimentally observed deposition patterns for two specific crevice geometries at a single flow rate. However, the robustness of the numerical model had not been evaluated with respect to alterations in flow rate or surface chemistry. Therefore, the purpose of this study was to extend the simulations of Wu et al. (2017) to explore the influence of the surface chemistry and shear rate upon thrombosis in micro crevices, which we hypothesize reflects the equilibrium between platelet deposition and clearance.

## 2. Methods

The mathematical model of thrombosis consists of equations of motion for blood that determine the pressure and velocity fields, supplemented with a set of coupled convection-diffusion-reaction (CDR) equations that govern the transport and inter-conversion of chemical and biological species – both within the thrombus and in the free stream. The details of the model have been published elsewhere (Wu et al., 2017) and are briefly summarized below.

### 2.1. Equations of motion for blood

Blood is treated as a multi-constituent mixture comprised of a fluid component, modeled as a linear (Newtonian) viscous fluid, and a thrombus component, treated as a porous medium. The fluid component, which accounts for the red blood cells (RBCs) and plasma, is governed by the equations of conservation of mass and linear momentum:

$$\frac{\partial \rho}{\partial t} + div(\rho \boldsymbol{v}) = 0 \quad (1)$$

$$\rho \frac{D\boldsymbol{v}}{Dt} = div\,\boldsymbol{T} + \rho \boldsymbol{b} - C_2 f(\phi)(\boldsymbol{v} - \boldsymbol{v}_T) \quad (2)$$

where $\rho$ is the density, $\boldsymbol{v}$ and $\boldsymbol{v}_T$ are the velocity of the fluid component and the thrombus component, respectively. The scalar field $\phi$ represents the local volume fraction of the deposited platelets (thrombus). Since the thrombus is stationary relative to its substrate, $\boldsymbol{v}_T$ represents the velocity of the biomaterial surface (in this study $\boldsymbol{v}_T = 0$); $\boldsymbol{b}$ is the body force; $\boldsymbol{T}$ is the Cauchy stress tensor of the fluid component, represented by:

$$\boldsymbol{T} = [-p(1-\phi)]\boldsymbol{I} + 2\mu(1-\phi)\boldsymbol{D} \quad (3)$$

where $p$ is the pressure, $\boldsymbol{I}$ is the identity tensor, $\mu$ is the asymptotic viscosity (3.5cP) of blood, $\boldsymbol{D}$ is the symmetric part of the velocity gradient tensor, $\boldsymbol{D} = {}^1/_2 [(grad\,\boldsymbol{v}) + (grad\,\boldsymbol{v})^T]$.

The density of the fluid component is given in terms of the volume fraction according to, $\rho = (1 - \phi)\rho_0$, where $\rho_0$ is the density of the fluid component (1060 kg/m$^3$) in the reference configuration, i.e., prior to any thrombus formation.

Free-flowing platelets are treated as a disperse medium that does not affect the physics of the flow. A growing thrombus, however, alters the local hemodynamics by means of a resistance force on the fluid, given by the last term on the right hand side of Eq. (2). The coefficient $C_2 = 2 \times 10^9 kg/(m^3 s)$ is computed assuming that the deposited platelets behave like densely compact particles (2.78um in diameter), described by Johnson et al. (1991) and Wu et al. (2014a, 2014b).

*2.2. Convection-diffusion-reaction equations for thrombosis*
The thrombosis model includes seven fundamental mechanisms involved in coagulation: platelet activation, platelet deposition, thrombus propagation, thrombus erosion by shear cleaning, thrombus stabilization, thrombus inhibition and thrombus-fluid interaction. (See Fig. 2) The model considers five categories (states) of platelets and five biochemical species, summarized in Table 1.

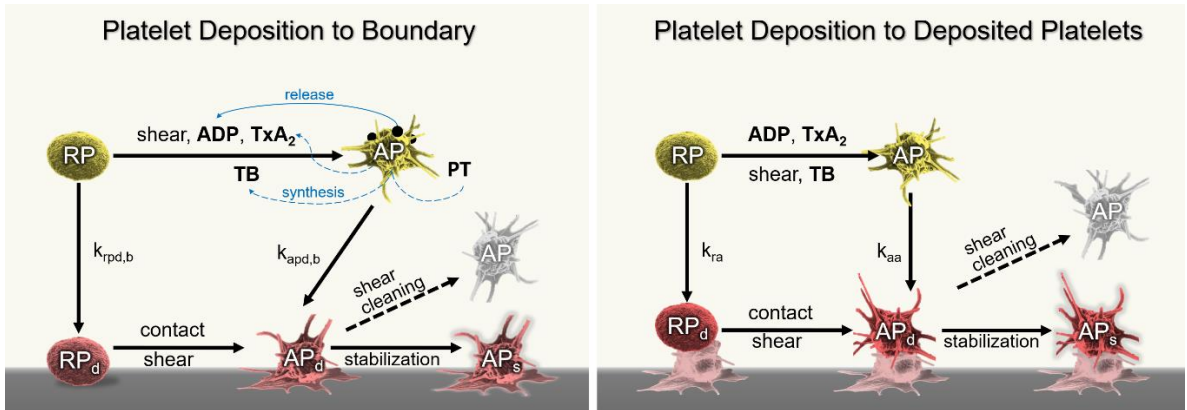

**Fig. 2.** Schematic depiction of the thrombosis model, comprised of platelet activation, deposition, aggregation, stabilization, and shear cleaning. The constants $k_{pa}$, $k_{ra}$, $k_{aa}$, $k_{rpd,b}$, $k_{apd,b}$, refer to the reaction rates for inter-conversion of the associated platelet states, where the suffix *b* refers to the reaction with the boundary (surface). (Adapted from Wu et al., 2017.)

**Table 1.** Categories of platelet species and biochemical species in the model.

| Platelet species | | Biochemical species | |
|---|---|---|---|
| Model abbreviation, $[C_i]$ | Description | Model abbreviation, $[C_i]$ | Description |
| RP | Resting (unactivated) PLTs | $a_{pr}$ | PLT-released agonists (ADP) |
| AP | Activated PLTs | $a_{ps}$ | PLT-synthesized agonists (TxA2) |
| $RP_d$ | Deposited Resting PLTs | TB | Thrombin |
| $AP_d$ | Deposited Activated PLTs | PT | Prothrombin |
| $AP_s$ | Deposited and Stabilized PLTs | AT | Antithrombin (ATIII) |

[RP], the local concentration of unactivated platelets in the flow field, can be activated and converted into [AP] via two mechanisms. Biochemical activation occurs when the aggregate concentration of agonists [$a_{pr}$] (ADP), [$a_{ps}$] (TxA2), and [TB] (thrombin) exceeds a critical value. The criterion for mechanical shear activation is adopted from Goodman et al. (2005) and Hellums (1994). Upon activation, platelets release ADP and continue synthesizing TxA2 in the activated state. Thrombin is synthesized on the activated platelet phospholipid membrane from prothrombin, [PT], and is inhibited by anti-thrombin III, [AT], mediated by heparin. Free-flowing platelets can deposit onto boundaries (surfaces) or on previously deposited platelets – [$RP_d$], [$AP_d$], [$AP_s$] – constituting a platelet clot. Activated platelets [AP] deposit at a greater rate than [RP]. For both states, the rates of deposition are substrate-specific. Shear stress exerted by the fluid component can clear deposited platelets from surfaces and erode the clot periphery.

The transport of the species in the flow field is described by a set of convection-diffusion-reaction equations:

$$\frac{\partial [C_i]}{\partial t} + div(\boldsymbol{v}[C_i]) = div(D_i \nabla [C_i]) + S_i \quad (4)$$

where $[C_i]$ is the concentration of species $i$; $D_i$ is the diffusivity of species $i$ in blood; and $S_i$ is the reaction source term for species $i$. Note that the subscript $i$ represents the species listed in Table 1. The deposition/cleaning of platelets is governed by concentration rate equations, which are simplified from Eq. (4) in the absence of convective and diffusive terms for [$RP_d$], [$AP_d$], and [$AP_s$]:

$$\frac{\partial [C_i]}{\partial t} = S_i \quad (5)$$

Reactions at a boundary, including deposition of platelets to surfaces, are modeled by surface-flux boundary conditions, following the approach of Sorensen et al. (1999). Further detailed mathematical description of the thrombosis model, including the specific values and expressions for all the source terms and their parameters, can be found in Wu et al. (2017).

*2.3. Computational domain and boundary conditions*

The simulations correspond to the experiments by Jamiolkowski et al. (2016) where human blood was perfused through a parallel-plate flow chamber with a defined micro-crevice. The computational domain only includes a portion of the overall 8-mm length of the experimental channel (Fig. 3(a)). The half-height of the main channel was 1.5 mm, and the depth was 0.1 mm. The side wall featured a rectangular micro-crevice, with height of 122 μm, and length, $L_c$, of either 53 μm, 90 μm, or 137 μm.

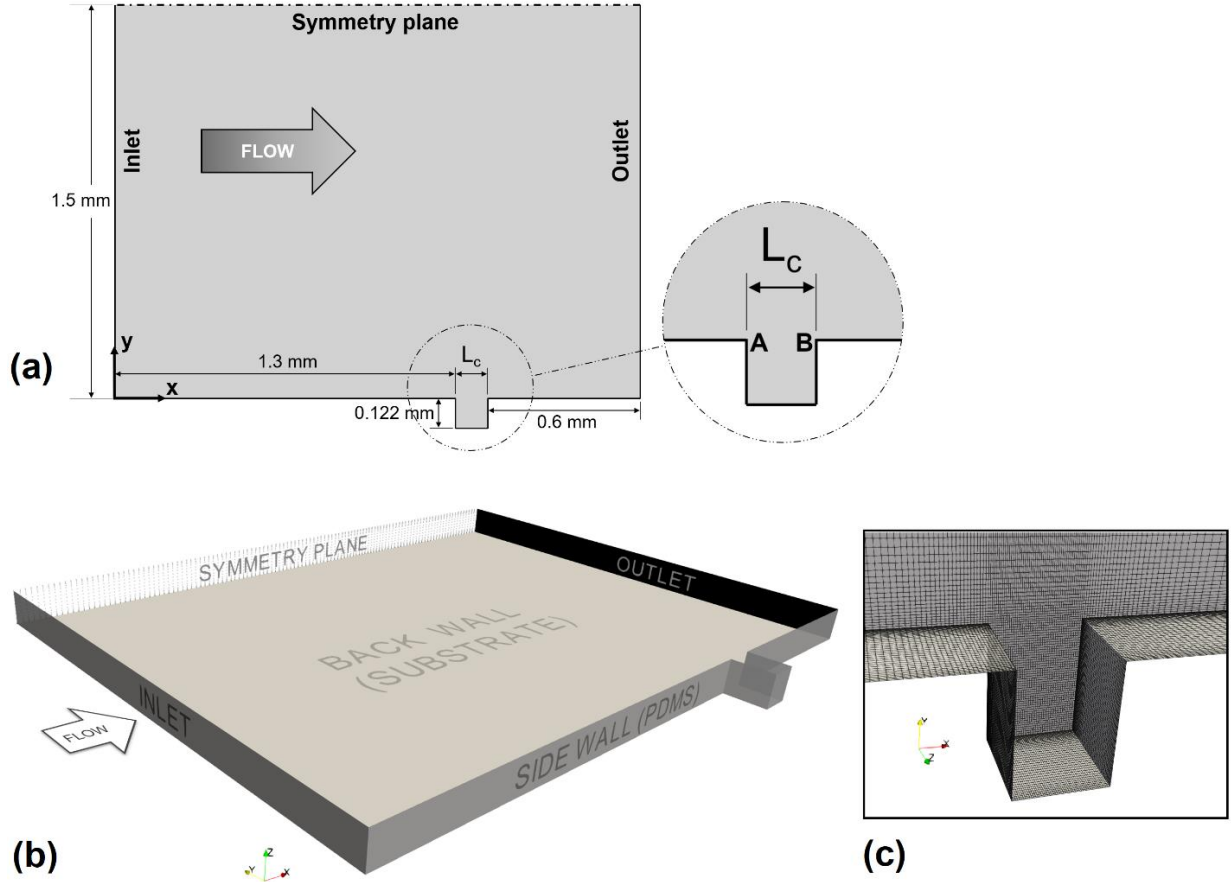

**Fig. 3.** Schematic of the computational domain. (a) Dimensions of the computational domain. The depth of the channel in z-direction is 0.1 mm, the length of the crevice, $L_c$, is 53 μm, 90 μm, or 137 μm. Inset: A and B label the upstream and downstream corners of the crevice, respectively. (b) Computational domain boundaries, annotated. The side wall and front wall (not shown) surfaces were made of PDMS, which is assumed passive in these simulations. The back wall surface was the substrate for platelet deposition: Ti6Al4V, collagen-coated glass (positive control) or MPC-Ti6Al4V (negative control). (c) Computation mesh showing the crevice region.

The top and side walls of the channel were made of polydimethylsiloxane (PDMS) and were passivated with bovine serum albumin (BSA) prior to blood perfusion. A titanium alloy sample (Ti6Al4V) acted as the back plate of the chamber and presented a substrate for platelet deposition. The experiments with the 90-μm crevice were repeated with a collagen-coated glass coverslip acting as a positive-control thrombogenic surface. For a negative control surface, a Ti6Al4V surface coated with methacryloyloxyethyl phosphorylcholine polymer (MPC-Ti6Al4V) was employed (Ye et al., 2010).

The computational domain used in this study is shown in Fig. 3(b). In the numerical simulations, the PDMS surfaces were assumed non-reactive, i.e., the boundary reaction rates were set to zero. The back wall, in contrast, supports platelet adhesion with the boundary reaction rates calibrated according to the surface chemistry. A fixed mass flow rate was prescribed at the inlet, with flow rate of 0.125 mL/min corresponding to wall shear rate of 400 $s^{-1}$ (Re = 0.4), and 0.310 mL/min corresponding to 1000 $s^{-1}$ (Re = 1.0). These wall shear rates are representative of regions within VADs that often contain undesired steps and crevices: inlet and outlet connectors, upstream and downstream of the stators in axial VADs, and outer wall of centrifugal VADs (Song et al., 2004; Zhang et al., 2008). Platelets were introduced at the inlet as uniform concentrations of [RP] and [AP], prescribed at $2.65 \times 10^{14}$ PLTs/m$^3$ and $5 \times 10^{12}$ PLTs/m$^3$, respectively, matching the experimental conditions. Simulations were performed for all 10 combinations of crevice sizes, biomaterial surfaces, and wall shear rates reported by Jamiolkowski et al. (2016). (See Table 2.) The mathematical model was numerically implemented in an open-source finite volume software library OpenFOAM (The OpenFOAM Foundation Ltd). Hexahedral meshes containing 230,000, 400,000, and 600,000 cells were created using OpenFOAM's blockMesh utility, and the 400,000-cell mesh (Fig. 3(c)) was selected based on mesh independence. Time step was 0.01 s and each thrombosis simulation required approximately 8 hours using 40 CPUs (2.4 GHz). The simulations were performed using OpenFOAM v6 and post-processing was done in ParaView 5.6.

**Table 2.** Simulation matrix for three crevice sizes and three biomaterials at two shear rates.

| Crevice width, $L_c$ (μm) | Biomaterial | Shear rate (s$^{-1}$) |
|---|---|---|
| 53 | Ti6Al4V | 400 |
|  |  | 1000 |
| 90 | Ti6Al4V | 400 |
|  |  | 1000 |
| 137 | Ti6Al4V | 400 |
|  |  | 1000 |
| 90 | Collagen | 400 |
|  |  | 1000 |
| 90 | MPC-Ti6Al4V | 400 |
|  |  | 1000 |

## 3. Results
*3.1. Flow field and transport of species*
Fig. 4 – Fig. 6 show the simulation results for a representative case: 90-μm crevice with Ti6Al4V at the high shear condition, 1000 s$^{-1}$. The velocity field on the middle plane of the channel (z-coordinate of 0.05 mm) at the beginning of the simulation is illustrated in Fig. 4(a). A large vortex is visible that forms inside the crevice. (The length of the vector arrows is fixed, and color scale corresponds to the velocity magnitude.) Fig. 4(b) shows the wall shear rate on the back and side walls of the channel. (Note the log-scale.) The shear rate approaches 1000 s$^{-1}$ in the main flow, drops by an order of magnitude inside the crevice, and is nearly zero in the bottom of the crevice. However, there is another prominent stagnation region near the main channel, connecting the two top corners of the crevice (denoted by letters A and B in the Fig. 3(a) inset). This arc of near-zero shear is formed where the main channel flow meets the counter rotating secondary flow. This is more readily visible in Fig. 4(c) illustrating velocity vectors and streamlines in the vicinity of the back wall. Notice the out-of-plane component of the velocity field (highlighted in Fig. 4(d)), originating from the upstream corner and moving towards the mid-plane, before reversing direction and impinging on the back wall.

Fig. 5 presents the thrombosis simulation results after 600 seconds for the above case, compared to the corresponding experimental data. A three-dimensional rendering of the thrombus, comprised of deposited platelets, is shown in Fig. 5(a). Solid red represents the region where volume fraction of deposited platelets, $\phi$, exceeds 0.8; translucent red represents $\phi$ from 0.5 to 0.8. Note that the clots are adherent to the back wall in both experiment and simulation. The simulated thrombus growth pattern closely matches the deposition observed experimentally by Jamiolkowski et al. (2016), shown in Fig. 5(b), where platelets accumulated at the two top corners of the crevice, with the clot at the downstream corner extending towards the bottom of the crevice. In the simulation, the deposition at the top corners follows the shape of the stagnation region shown in Fig. 4(b), whereas the downward-extending clot corresponds to the region of impinging flow highlighted in Fig. 4(d). The simulation results in Fig. 5(c)-(d) demonstrate the effect of the growing thrombus on the flow field inside and near the crevice. The flow around the clot is disturbed due to the thrombus resistance term in Eq. (2), which alters the platelet transport and shapes further growth. On the other hand, the elevated shear on the outer surface of the clot opposes the growth by shear cleaning of platelets.

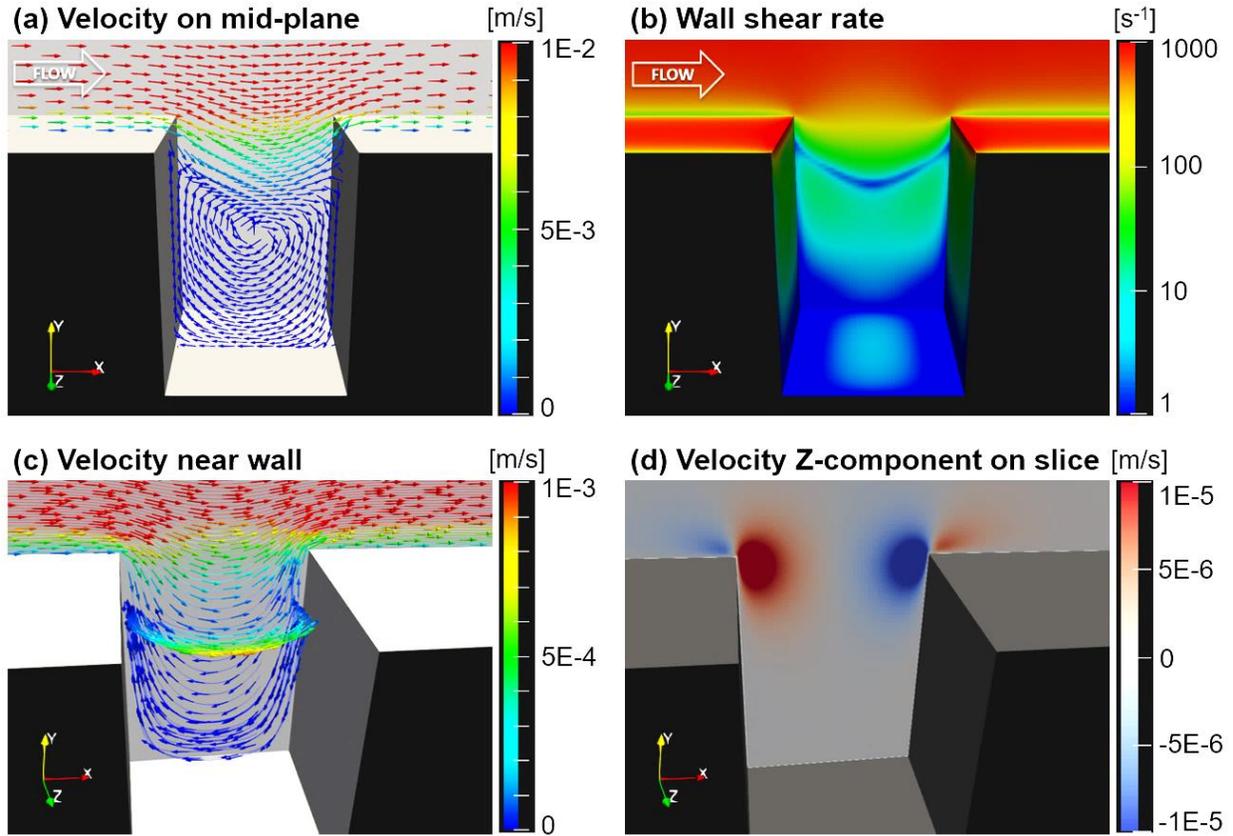

**Fig. 4.** Flow field in the 90-μm crevice with Ti6Al4V at the high shear rate condition (1000 $s^{-1}$) prior to thrombus formation. (a) Velocity vectors on the middle plane of the channel (0.05 mm from the back wall). Length of the vector arrows is fixed, and color scale corresponds to the velocity magnitude. (b) Log wall shear rate on the back and side walls. Note the stagnation region (near-zero shear) connecting the top corners of the crevice. (c) Velocity streamlines and vectors visualized near the back wall, colored by velocity magnitude. (d) Z-component of the velocity (normal to the back wall), shown on a plane 5 μm from the back wall, reveals the flow directed away from and towards the substrate surface.

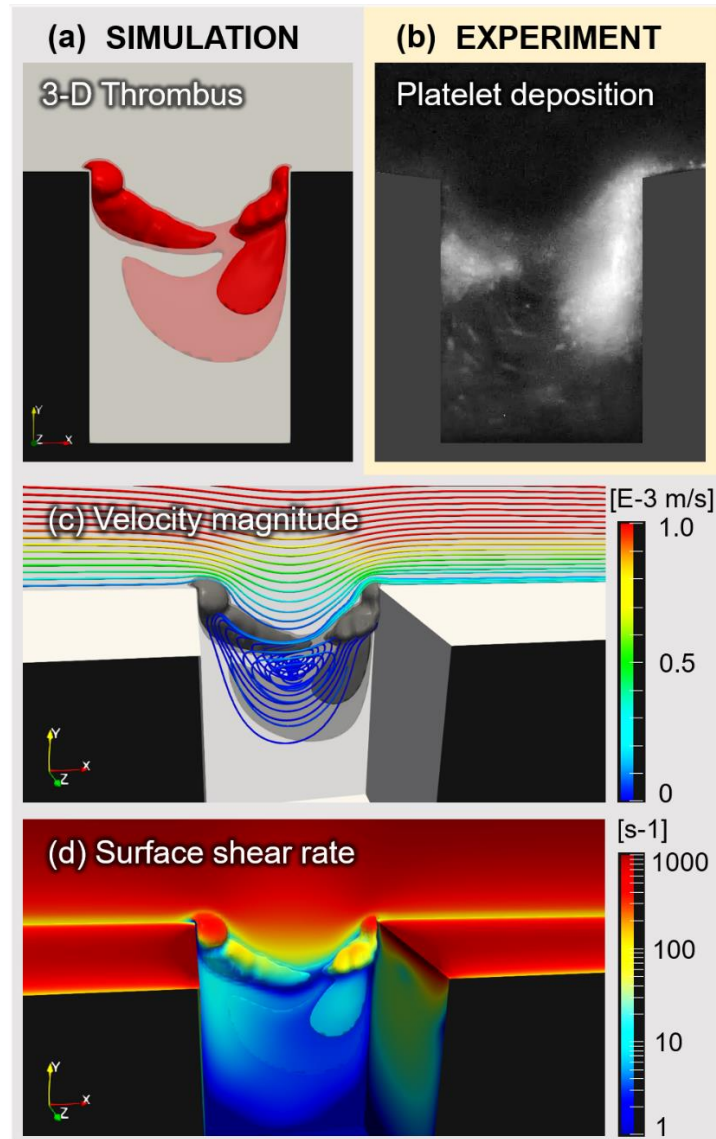

**Fig. 5.** Simulations results in the 90-μm crevice with Ti6Al4V at the high shear rate condition ($1000\ s^{-1}$) after 600 seconds compared to the experimental results. Flow is from left to right. (a) 3D rendering of the simulated thrombus. Solid red represents region in space where volume fraction of deposited platelets, $\phi$, exceeds 0.8; translucent red represents $\phi$ from 0.5 to 0.8. (b) Platelet deposition observed experimentally by Jamiolkowski et al. (2016) using fluorescent platelets. Bright regions correspond to accumulation of deposited platelets. (c) Streamlines distorted by thrombi. (d) Surface shear rate on the walls and the thrombus surface shown in log-scale.

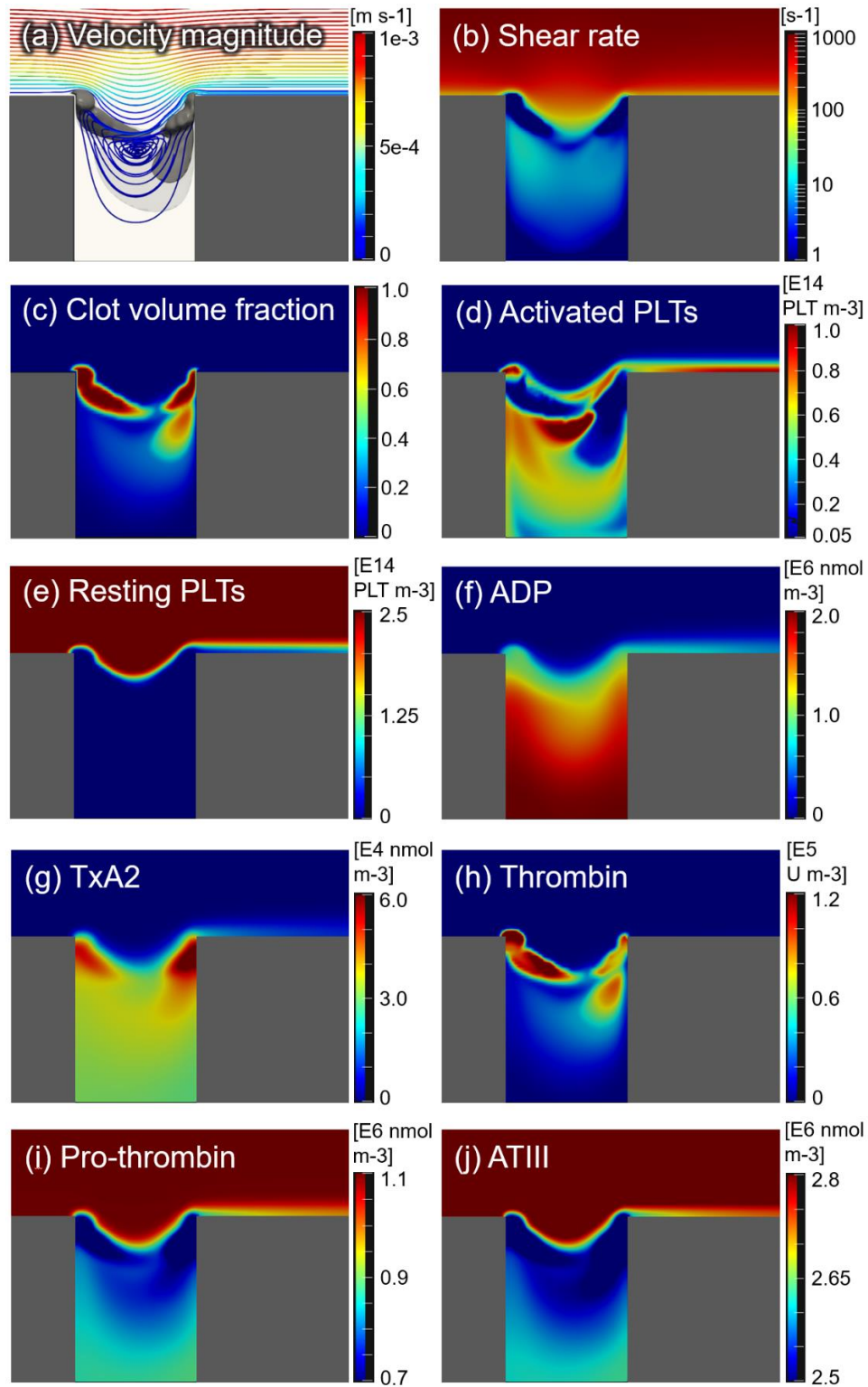

**Fig. 6.** Flow around thrombus and concentration fields of platelet and biochemical species near back wall (z = 2 µm). Results shown for the 90-µm crevice in Ti6Al4V at the high shear rate condition (1000 s$^{-1}$) after 600 seconds of simulation. The color scale selected to highlight the local variations in concentration fields.

Fig. 6 provides a collage of the flow field (Fig. 6(a)-(b)) and platelet/biochemical species (Fig. 6(c)-(j)) at the simulation timepoint above. The volume fraction of deposited platelets is greatest in the thrombus core, as shown in Fig. 6(c). The free-flowing platelets, [AP] and [RP], shown in Fig. 6(d)-(e), are excluded from the space occupied by the thrombus. There is an excess of [AP] in the center of the vortex visible in Fig. 6(a). Fig. 6(f)-(h) show the concentration fields of agonists: ADP, TxA2, and thrombin. ADP released by platelets upon activation accumulates within the crevice recirculation zone. TxA2 and thrombin are continuously synthesized by activated platelets and show peak concentrations in the interior of the thrombus, as well as elevated concentrations within the crevice compared to the main channel. Pro-thrombin and anti-thrombin, shown in Fig. 6(i)-(j), are consumed within the crevice and inside the thrombus. Note the tail of elevated [AP] and agonist concentrations extending downstream of the crevice.

*3.2. Effect of crevice width on thrombus deposition patterns.*
Simulated deposition patterns in Ti6Al4V at time = 600 s for all six combinations of crevice width, $L_c$, and shear rate are summarized in Fig. 7 (first column), juxtaposed with corresponding experimental microscopic image with fluorescently labeled human platelets (second column), as well as probability map of platelet deposition (third column) from Jamiolkowski et al. (2016). (The N/A frames refer to cases where raw data was no longer available at the time of publication.) In all cases, deposited platelets accumulated on the back wall at both upstream and downstream top corners of the crevice with no significant deposition in the depth of the crevice. The clot at the downstream corner of the smaller crevices ($L_c$ = 53 μm and 90 μm) is elongated in the vertical direction. This zone corresponds to the region of impinging flow highlighted in Fig. 4(d).

In all crevice sizes, the increase in flow rate results in greater platelet deposition. (Peclet number, which is ratio of convective transport to diffusive transport, $Pe \gg 1$.) The higher flow rate also shifts the location of the stagnation line and the associated platelet deposition further down the back wall of the crevice. A clot extending outside the crevice into the main channel can be seen in $L_c$ = 53 μm and 90 μm under the low flow rate condition. However, this trailing clot is absent when the flow rate is increased. Supplementary video showing the time progression of the thrombus growth can be found in the online version of this article.

Fig. 8 shows the thrombus volume plotted against simulation time for the three crevices sizes at two flow rates. The narrowest crevice with $L_c$ = 53 μm shows the greatest amount of deposition compared to the other two under the same flow rate. The widest crevice, despite being 260% larger in volume, accumulated the least amount of thrombus. The higher flow rate results in a greater thrombus volume in all three crevices, but the effect is less pronounced in $L_c$ = 137 μm. The percent increase in clot volume at 600 seconds between the low and high flow rates is 106% for $L_c$ = 53 μm, 97% for $L_c$ = 90 μm, but only 19% for $L_c$ = 137 μm.

Towards the end of the simulation, the combined concentration of agonists (ADP, TxA2, and thrombin) exceeds the platelet activation threshold in $L_c$ = 53 μm and $L_c$ = 90 μm but does not reach that level in $L_c$ = 137 μm. To quantify the accumulation of agonists and subsequent increase in activated platelet concentration within the crevices, the volume where local [AP] exceeds 50% of the total global platelet concentration was calculated. At 600 seconds in the high-flow case, local [AP] concentration exceeded $1.35 \times 10^{14}$ PLT/m$^3$ in 90% of the 53-μm crevice volume, in 80% of the 90-μm crevice volume, but only 0.6% of the 137-μm crevice volume.

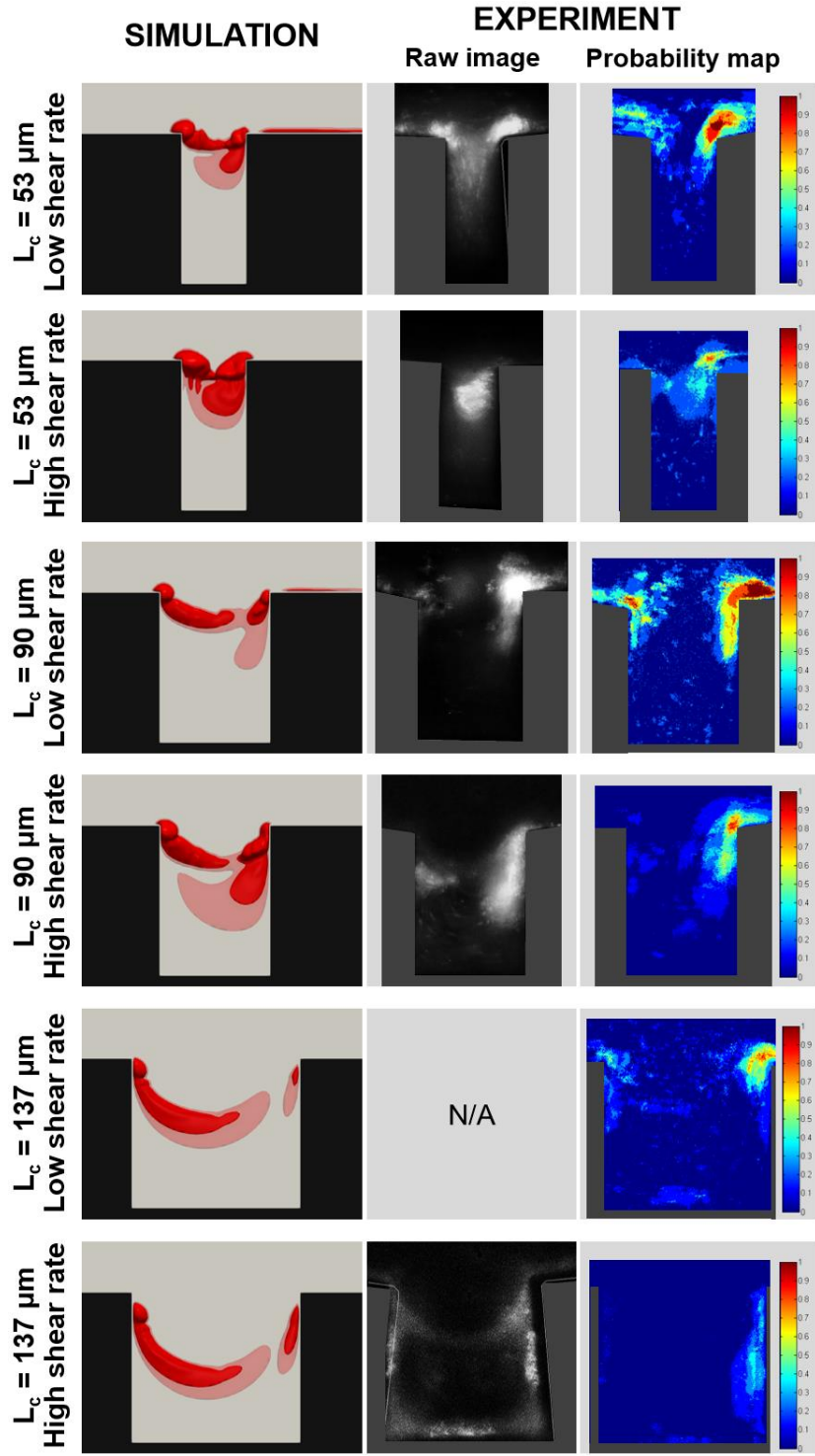

**Fig. 7.** The effect of crevice width, $L_c$, and shear rate on simulated deposition patterns in Ti6Al4V. Simulations results at time = 600 s (first column) compared to experimental data: microscopic images with fluorescently labeled human platelets, where bright regions correspond to accumulation of deposited platelets (second column), and probability maps of platelet deposition (third column). Experimental data from Jamiolkowski et al. (2016) reprinted with permission from Elsevier.

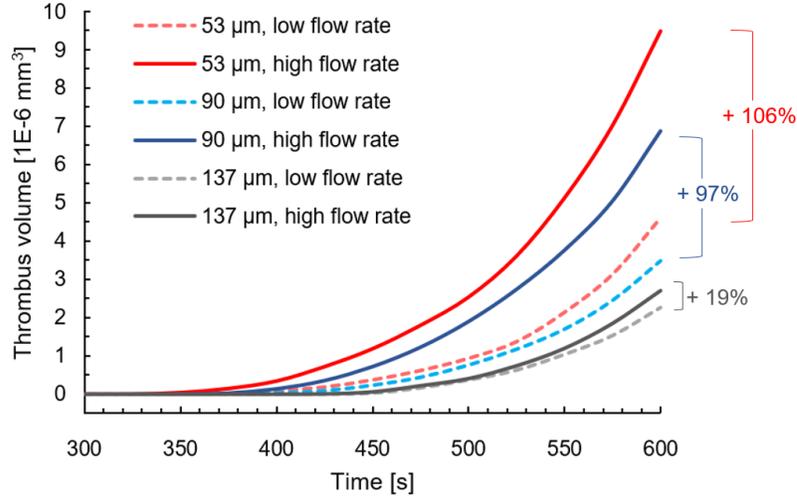

**Fig. 8.** The effect of crevice width, $L_c$, and shear rate on the simulated thrombus growth in Ti6Al4V. Only the computational cells with the volume fraction of deposited platelets $\phi \geq 0.8$ are included in the thrombus volume. Prior to the 300-second mark, the deposited platelets accumulate in the boundary field (surface), which is not reflected in the plots.

### 3.3. Effect of surface chemistry on thrombus deposition patterns.

The effect of surface chemistry was simulated by prescribing the boundary reaction rates for [AP] and [RP] specific to each substrate (back wall), summarized in Table 3. The reaction constants for [RP] were selected assuming resting unactivated platelets show virtually no deposition on Ti6Al4V and MPC-Ti6Al4V surfaces but have a greater affinity to the collagen-coated surface. [AP] had significantly greater reaction rates on all materials, signifying that the thrombosis was primarily driven by the activated platelet deposition. The positive and negative control surfaces (collagen and MPC-Ti6Al4V, respectively) were separated by an order of magnitude difference in [AP] reaction rate. The rest of the parameters were the same as in previous simulations of Wu et al. (2017).

Simulated platelet deposition on three different substrates at time = 600 s for all six combinations of surface chemistry and shear rate are summarized in Fig. 9, juxtaposed with corresponding experimental microscopic image with fluorescently labeled human platelets, as well as probability map of platelet deposition from Jamiolkowski et al. (2016). The crevice width $L_c = 90$ μm was the same in all cases. Compared with Ti6Al4V described in the previous section, the deposition on the collagen substrate was significantly exacerbated. Large thrombi formed along the entire length of the crevice on the collagen substrate, similar to experimental observations. In addition, there was substantial deposition in the main channel outside the crevice. In contrast, MPC-Ti6Al4V showed the least amount of deposition, limited to the top corners of the crevice.

**Table 3.** Substrate-specific boundary reaction rates for the three materials, prescribed at the back wall of the crevice.

|  | Ti6Al4V | Collagen | MPC-Ti6Al4V |
|---|---|---|---|
| RP-boundary(wall) deposition rate, $k_{rpd,b}$ (m/s) | $1.0 \times 10^{-20}$ | $1.0 \times 10^{-8}$ | $1.0 \times 10^{-20}$ |
| AP-boundary(wall) deposition rate, $k_{apd,b}$ (m/s) | $1.0 \times 10^{-6}$ | $3.0 \times 10^{-6}$ | $3.0 \times 10^{-7}$ |

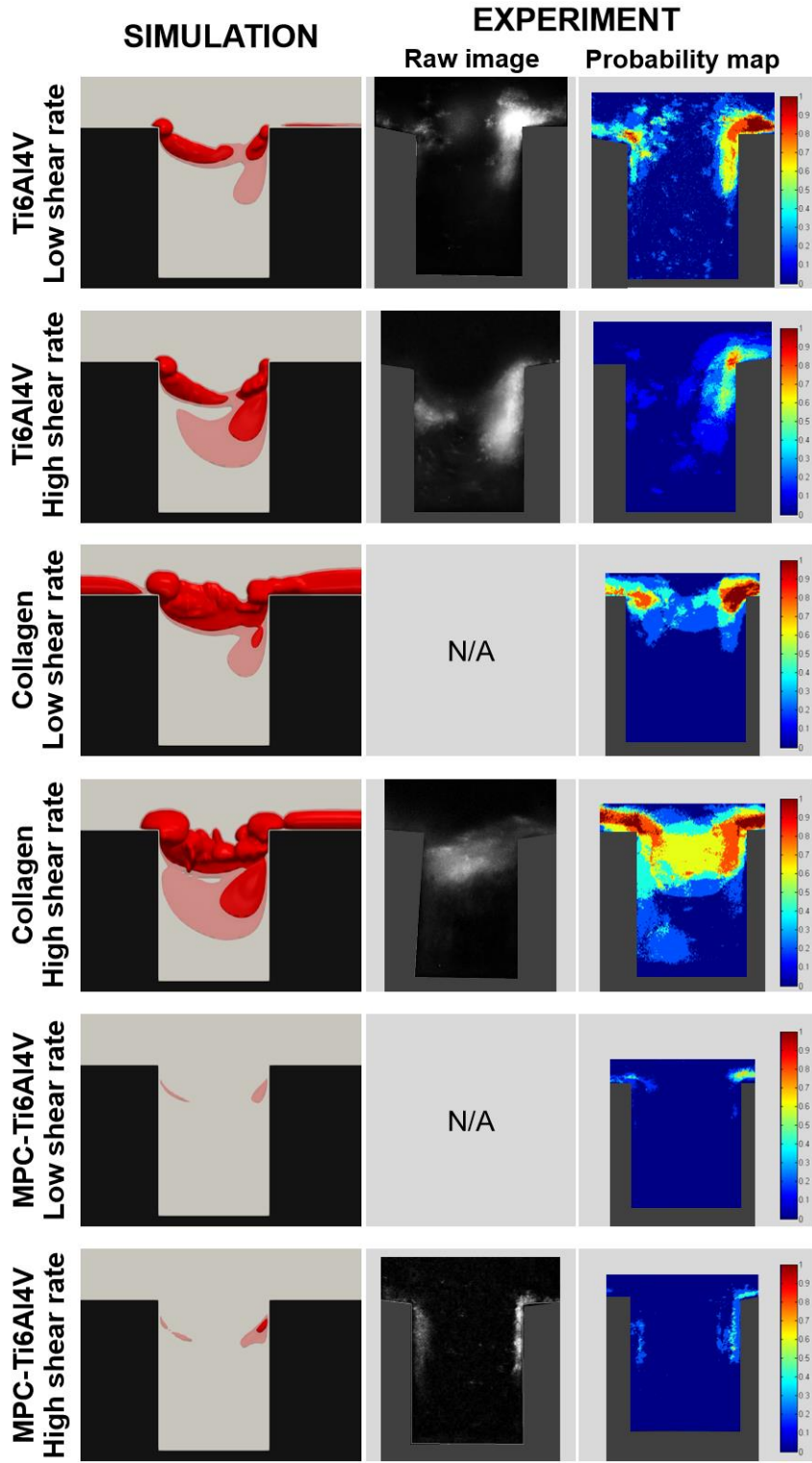

**Fig. 9.** The effect of surface chemistry and shear rate on simulated deposition patterns in the crevice with $L_c = 90$ μm. Simulations results at time = 600 s (first column) compared to experimental data: microscopic images with fluorescently labeled human platelets, where bright regions correspond to accumulation of deposited platelets (second column), and probability maps of platelet deposition (third column). Experimental data from Jamiolkowski et al. (2016) reprinted with permission from Elsevier.

With all materials, the deposition inside the crevice increased in size at the greater flow rate. The opposite trend was observed outside the crevice: on collagen, the extended clot upstream of the crevice is present at the low flow rate but disappears at the high flow rate. Similarly, Ti6Al4V exhibits a clot downstream of the crevice under the low flow condition but not the high flow condition.

## 4. Discussion

Thrombosis in mechanical circulatory support devices is a balance of many factors (or a disruption thereof) including advection and deposition of platelets as well as removal by shear cleaning (Lowe, 2003; Selmi et al., 2019). These are influenced by the characteristics of the flow which is a function of the topology/geometry (size/shape) and by surface chemistry of the biomaterial. This study explored this balance in the setting of blood flow over a micro-crevice. The counter-intuitive location of the clots at the top corners of the crevice (Fig. 5) emerges from the complex flow within this geometry. The deposition at the upstream corner is initiated when the incoming activated platelets encounter the stagnation zone formed at the border between the main flow and the vortical flow inside the crevice, illustrated in Fig. 4. Since the wall shear rate in the stagnation zone is nearly zero, the depositing platelets are not cleaned by shear stress. The deposition at the downstream corner is further enhanced by the three-dimensional flow structure highlighted in Fig. 4(d), where the significant secondary component of the flow (out of/into viewing plane) results in impingement of platelets onto the back wall. This clot then propagates in the direction of the vortical flow, down and towards the center of the crevice.

A larger crevice is less stagnant and therefore more readily washed. A deep and short crevice is more susceptible to accumulation of agonists and growth of thrombus. A prominent vortex was present in the 53-μm and 90-μm wide crevices but not in the 137-μm wide crevice. Fig. 6 reveals the build-up of agonists within the depth of the 90-μm crevice, corresponding to the recirculation bubble; hence platelets in this region are exposed to a greater concentration of agonists (ADP, TxA2, and thrombin), and for a greater time. This positive feedback loop results in an explosive increase in platelet activation. In contrast, the combined concentration of agonists did not exceed the activation threshold in the 137-um wide crevice after 600 seconds.

We see in Fig. 7 through Fig. 9 that the effect of increasing the flow rate is two-fold: it increases the incoming supply of activated platelets but also affects removal/clearance due to shear cleaning. This is demonstrated in Fig. 9 with Ti6Al4V and collagen cases, wherein the greater flow rate resulted in greater deposition within the crevice but reduced the deposition outside the crevice. Spatial variation in the shear field offers an explanation (see Fig. 5(d)). Increasing the flow rate by a factor of 2.5 proportionally increases the wall shear rate in the main channel, but the effect on shear levels within the crevice is inconsequential in face of the two-orders-of-magnitude drop. Since the increased platelet transport at higher flow is not contested by shear cleaning within the crevice, the resulting clot is larger. However, shear cleaning prevails outside the crevice. The absence of a prominent recirculation zone (and associated diminution in shear rate) in the widest crevice also explains the more pronounced effect of flow rate on clot growth in the thinner crevices versus the almost negligible effect in the widest crevice (Fig. 8). With respect to surface chemistry, a more reactive surface such as collagen is able to capture incoming platelets at a greater rate, tilting the balance between deposition and cleaning in favor of thrombosis, while a highly hemocompatible surface like MPC-Ti6Al4V can suppress deposition (Fig. 9).

The simulation results, supplementing the experimental results of Jamiolkowski et al. (2016) that prompted this study, offer insights for design and operation of blood-contacting devices. Increasing the flow rate can reduce platelet deposition on well-washed surfaces but might exacerbate deposition in low-shear zones such as crevices. In addition, if crevices cannot be avoided in a blood-contacting device, tailoring the crevice size/width to the shear conditions can help promote washout and reduce the risk of thrombosis. Use of highly biocompatible coatings such as MPC-polymer can further reduce platelet deposition. Durability of such coatings under high shear rates can be a concern (Sin et al., 2009). Our results suggest that using Ti6Al4V in the main flow path and applying MPC coating locally in the low-shear recirculation zones could be a viable strategy.

Some limitations of this study are worth noting. The simulations assumed the side and front walls of the channel (made of PDMS in the experiment) as completely nonreactive to isolate the effect of the substrate used at the back wall. This introduced a level of disparity between simulation and experiment due to non-negligible reactivity of the PDMS surfaces. Jamiolkowski et al. (2016) reported platelets deposition on PDMS side walls of the 137-µm crevice at high shear rate, seen in Fig. 7 raw image. Although using the substrate material on the side walls would have been desirable, the use of PDMS was partly motivated by pragmatic reasons related to construction of the devices. We should mention that the raw images presented in Fig. 7 and Fig. 9 are snapshots from a video and the fluorescent signal does not discriminate between deposited platelets and recirculating platelets in motion. However, the bright regions in the images correspond to accumulation of static deposited platelets. The reader is directed to videos included in the supplementary material of Jamiolkowski et al. (2016) to appreciate the dynamics of platelet motion. We did not account explicitly for near-wall platelet excess due to red blood cell (RBC) trafficking (Crowl and Fogelson, 2011; Fåhræus and Lindqvist, 1931; Fitzgibbon et al., 2015; Karino and Goldsmith, 1979; Spann et al., 2016; Zhao et al., 2008, 2007). Since our thrombosis model uses a continuum approach (Fogelson, 1992; Goodman et al., 2005; Sorensen et al., 1999; Wu et al., 2017) it could be supplemented with a multiphase continuum representation (Wu et al., 2014; 2014a; 2020) in combination with extended diffusion model (Hund and Antaki, 2009) to include RBC trafficking and platelet margination. We also note that this work investigates a low Reynolds number scenario, so the near-wall flow environment could be different in full-scale devices. Additionally, although the current model accurately captures the balance between platelet deposition and shear cleaning, it cannot simulate the stochastic processes of platelet translocation, clot deformation, and embolization – phenomena reproduced in particle-based models (Flamm et al., 2012; Qi et al., 2019; Wang et al., 2020; Yazdani et al., 2017; Zhang et al., 2014) and phase-field models (Xu et al., 2017; Zheng et al., 2020). The shear-induced platelet activation mechanism adopted in this work lacks a more detailed treatment of models that consider full shear accumulation history (Soares et al., 2013) and contributions of different frequency components of shear stress (Consolo et al., 2017). We also did not include the effect of von Willebrand Factor (vWF) on platelet deposition, which could be significant in the high-shear zones outside the crevice (Casa et al., 2015; Mehrabadi et al., 2016; Ruggeri, 2009; Yazdani et al., 2017).

Ongoing work aims to address the above limitations and extends this study to clinical devices such as HeartMate II and HeartWare HVAD where micro-scale crevices are incorporated in a macro-scale device geometry. The current model development focuses on implementing the mechanism of vWF unfolding and subsequent vWF-mediated thrombosis in high-shear conditions. We are also conducting a global sensitivity analysis of the model to better characterize the influence of the various model parameters.

## 5. Conclusion
The multi-constituent simulations reproduced the platelet deposition patterns observed experimentally in micro crevices, and elucidated the role of shear, shear rate, and surface chemistry. The results offer insights for design and operation of blood-contacting devices.

**Declaration of Competing Interest**

The mention of commercial products and/or manufacturers does not imply endorsement by the FDA or the U.S. Department of Health and Human Services.

**Acknowledgments**

Author Wei-Tao Wu thanks the support of the grant NSFC 11802135. This work was supported by the National Institute of Health grant R01HL089456.

We would like to extend our gratitude to Dr. Francis D. Pagani (Michigan Medicine, University of Michigan) for providing the HM2 explant image.

**Supplementary material**

Supplementary video with simulation results showing the time progression of thrombus growth can be found in the online version of this article.

**References**

Acharya, D., Loyaga-Rendon, R., Morgan, C.J., Sands, K.A., Pamboukian, S. V., Rajapreyar, I., Holman, W.L., Kirklin, J.K., Tallaj, J.A., 2017. INTERMACS Analysis of Stroke During Support With Continuous-Flow Left Ventricular Assist Devices. JACC Hear. Fail. 5, 703–711. https://doi.org/10.1016/j.jchf.2017.06.014

Bhamidipati, C.M., Ailawadi, G., Bergin, J., Kern, J.A., 2010. Early thrombus in a HeartMate II left ventricular assist device: A potential cause of hemolysis and diagnostic dilemma. J. Thorac. Cardiovasc. Surg. 140, e7. https://doi.org/10.1016/j.jtcvs.2009.09.046

Bozzo, J., Tonda, R., Hernández, M.R., Alemany, M., Galán, A.M., Ordinas, A., Escolar, G., 2001. Comparison of the effects of human erythrocyte ghosts and intact erythrocytes on platelet interactions with subendothelium in flowing blood. Biorheology 38, 429–37.

Casa, L.D.C., Deaton, D.H., Ku, D.N., 2015. Role of high shear rate in thrombosis. J. Vasc. Surg. 61, 1068–1080. https://doi.org/10.1016/j.jvs.2014.12.050

Cho, S.-M., Hassett, C., Rice, C.J., Starling, R., Katzan, I., Uchino, K., 2019. What Causes LVAD-Associated Ischemic Stroke? Surgery, Pump Thrombosis, Antithrombotics, and Infection. ASAIO J. 65, 775–780. https://doi.org/10.1097/MAT.0000000000000901

Consolo, F., Sheriff, J., Gorla, S., Magri, N., Bluestein, D., Pappalardo, F., Slepian, M.J., Fiore, G.B., Redaelli, A., 2017. High Frequency Components of Hemodynamic Shear Stress Profiles are a Major Determinant of Shear-Mediated Platelet Activation in Therapeutic Blood Recirculating Devices. Sci. Rep. 7, 1–14. https://doi.org/10.1038/s41598-017-05130-5

Crowl, L., Fogelson, A.L., 2011. Analysis of mechanisms for platelet near-wall excess under arterial blood flow conditions. J. Fluid Mech. 676, 348–375. https://doi.org/10.1017/jfm.2011.54

Fåhræus, R., Lindqvist, T., 1931. THE VISCOSITY OF THE BLOOD IN NARROW CAPILLARY TUBES. Am. J. Physiol. Content 96, 562–568. https://doi.org/10.1152/ajplegacy.1931.96.3.562

Fitzgibbon, S., Spann, A.P., Qi, Q.M., Shaqfeh, E.S.G., 2015. In Vitro Measurement of Particle Margination in the Microchannel Flow: Effect of Varying Hematocrit. Biophys. J. 108, 2601–2608. https://doi.org/10.1016/j.bpj.2015.04.013

Flamm, M.H., Colace, T. V, Chatterjee, M.S., Jing, H., Zhou, S., Jaeger, D., Brass, L.F., Sinno, T.,

Diamond, S.L., 2012. Multiscale prediction of patient-specific platelet function under flow. Blood 120, 190–198.

Fogelson, A.L., 1992. Continuum Models of Platelet Aggregation: Formulation and Mechanical Properties. SIAM J. Appl. Math. 52, 1089–1110. https://doi.org/10.1137/0152064

Glass, C.H., Christakis, A., Fishbein, G.A., Watkins, J.C., Strickland, K.C., Mitchell, R.N., Padera, R.F., 2019. Thrombus on the inflow cannula of the HeartWare HVAD: an update. Cardiovasc. Pathol. 38, 14–20. https://doi.org/10.1016/j.carpath.2018.09.002

Goodman, P.D., Barlow, E.T., Crapo, P.M., Mohammad, S.F., Solen, K.A., 2005. Computational model of device-induced thrombosis and thromboembolism. Ann. Biomed. Eng. 33, 780–797.

Hellums, J.D., 1994. 1993 Whitaker Lecture: biorheology in thrombosis research. Ann. Biomed. Eng. 22, 445–455.

Hund, S.J., Antaki, J.F., 2009. An extended convection diffusion model for red blood cell-enhanced transport of thrombocytes and leukocytes. Phys. Med. Biol. 54, 6415–6435. https://doi.org/10.1088/0031-9155/54/20/024

Jamiolkowski, M.A., Pedersen, D.D., Wu, W., Antaki, J.F., Wagner, W.R., 2016. Visualization and analysis of biomaterial-centered thrombus formation within a defined crevice under flow. Biomaterials 96, 72–83. https://doi.org/10.1016/j.biomaterials.2016.04.022

Jamiolkowski, M.A., Woolley, J.R., Kameneva, M. V., Antaki, J.F., Wagner, W.R., 2015. Real time visualization and characterization of platelet deposition under flow onto clinically relevant opaque surfaces. J. Biomed. Mater. Res. Part A 103, 1303–1311. https://doi.org/10.1002/jbm.a.35202

Johnson, G., Massoudi, M., Rajagopal, K.R., 1991. Flow of a fluid—solid mixture between flat plates. Chem. Eng. Sci. 46, 1713–1723. https://doi.org/10.1016/0009-2509(91)87018-8

Karino, T., Goldsmith, H.L., 1979. Aggregation of human platelets in an annular vortex distal to a tubular expansion. Microvasc. Res. 17, 217–237. https://doi.org/10.1016/S0026-2862(79)80001-1

Kaufmann, F., Hörmandinger, C., Potapov, E., Krabatsch, T., Falk, V., 2018. HVAD Thrombosis - Searching for the Source. J. Hear. Lung Transplant. 37, S145–S146. https://doi.org/10.1016/j.healun.2018.01.352

Kirklin, J.K., Naftel, D.C., Pagani, F.D., Kormos, R.L., Stevenson, L.W., Blume, E.D., Myers, S.L., Miller, M.A., Baldwin, J.T., Young, J.B., 2015. Seventh INTERMACS annual report: 15,000 patients and counting. J. Hear. Lung Transplant. 34, 1495–1504. https://doi.org/10.1016/j.healun.2015.10.003

Kirklin, J.K., Pagani, F.D., Kormos, R.L., Stevenson, L.W., Blume, E.D., Myers, S.L., Miller, M.A., Baldwin, J.T., Young, J.B., Naftel, D.C., 2017. Eighth annual INTERMACS report : Special focus on framing the impact of adverse events. J. Hear. Lung Transplant. 36, 1080–1086. https://doi.org/10.1016/j.healun.2017.07.005

Kittipibul, V., Xanthopoulos, A., Hurst, T.E., Fukamachi, K., Blackstone, E.H., Soltesz, E., Starling, R.C., 2020. Clinical Courses of HeartMate II Left Ventricular Assist Device Thrombosis. ASAIO J. 66, 153–159. https://doi.org/10.1097/MAT.0000000000000952

Kormos, R.L., Cowger, J., Pagani, F.D., Teuteberg, J.J., Goldstein, D.J., Jacobs, J.P., Higgins, R.S., Stevenson, L.W., Stehlik, J., Atluri, P., Grady, K.L., Kirklin, J.K., 2019. The Society of Thoracic Surgeons Intermacs Database Annual Report: Evolving Indications, Outcomes, and Scientific Partnerships. Ann. Thorac. Surg. 107, 341–353. https://doi.org/10.1016/j.athoracsur.2018.11.011

Kreuziger, L.B., Slaughter, M.S., Sundareswaran, K., Mast, A.E., 2018. Clinical relevance of histopathologic analysis of heart mate II thrombi. ASAIO J. 64, 754–759. https://doi.org/10.1097/MAT.0000000000000759

Li, S., Beckman, J.A., Cheng, R., Ibeh, C., Creutzfeldt, C.J., Bjelkengren, J., Herrington, J., Stempien-Otero, A., Lin, S., Levy, W.C., Fishbein, D., Koomalsingh, K.J., Zimpfer, D., Slaughter, M.S., Aliseda, A., Tirschwell, D., Mahr, C., 2020. Comparison of Neurologic Event Rates Among HeartMate II, HeartMate 3, and HVAD. ASAIO J. 66, 620–624. https://doi.org/10.1097/MAT.0000000000001084

Li, S., Mahr, C., 2019. Evaluating ventricular assist device outcomes internationally with a focus on neurological events. Heart 105, 266 LP – 267. https://doi.org/10.1136/heartjnl-2018-313858

Lowe, G.D.O., 2003. Virchow's Triad Revisited: Abnormal Flow. Pathophysiol. Haemost. Thromb. 33, 455–457. https://doi.org/10.1159/000083845

Mcilvennan, C.K., Grady, K.L., Matlock, D.D., Helmkamp, L.J., Abshire, M., Allen, L.A., 2019. End of life for patients with left ventricular assist devices : Insights from INTERMACS. J. Hear. Lung Transplant. 38, 374–381. https://doi.org/10.1016/j.healun.2018.12.008

Mehrabadi, M., Casa, L.D.C., Aidun, C.K., Ku, D.N., 2016. A Predictive Model of High Shear Thrombus Growth. Ann. Biomed. Eng. 44, 2339–2350. https://doi.org/10.1007/s10439-016-1550-5

Mokadam, N.A., Andrus, S., Ungerleider, A., 2011. Thrombus formation in a HeartMate II. Eur. J. Cardiothorac. Surg. 39, 414. https://doi.org/10.1016/j.ejcts.2010.06.015

Najjar, S.S., Slaughter, M.S., Pagani, F.D., Starling, R.C., McGee, E.C., Eckman, P., Tatooles, A.J., Moazami, N., Kormos, R.L., Hathaway, D.R., Najarian, K.B., Bhat, G., Aaronson, K.D., Boyce, S.W., 2014. An analysis of pump thrombus events in patients in the HeartWare ADVANCE bridge to transplant and continued access protocol trial. J. Hear. Lung Transplant. 33, 23–34. https://doi.org/10.1016/j.healun.2013.12.001

Nguyen, A.B., Uriel, N., Adatya, S., 2016. New Challenges in the Treatment of Patients With Left Ventricular Support : LVAD Thrombosis. Curr. Heart Fail. Rep. 302–309. https://doi.org/10.1007/s11897-016-0310-z

Pappalardo, F., Bertoldi, L.F., Sanvito, F., Marini, C., Consolo, F., 2020. Inflow Cannula Obstruction of the HeartWare Left Ventricular Assist Device: What Do We Really Know? Cardiovasc. Pathol. 107299. https://doi.org/10.1016/j.carpath.2020.107299

Potapov, E., Kaufmann, F., Scandroglio, A.M., Pieri, M., 2017. Pump Thrombosis, in: Mechanical Circulatory Support in End-Stage Heart Failure. Springer International Publishing, Cham, pp. 495–512. https://doi.org/10.1007/978-3-319-43383-7_48

Qi, Q.M., Dunne, E., Oglesby, I., Schoen, I., Ricco, A.J., Kenny, D., Shaqfeh, E.S.G., 2019. In Vitro Measurement and Modeling of Platelet Adhesion on VWF-Coated Surfaces in Channel Flow. Biophys. J. 116, 1136–1151. https://doi.org/10.1016/j.bpj.2019.01.040

Rowlands, G.W., Pagani, F.D., Antaki, J.F., 2020. Classification of the Frequency, Severity, and Propagation of Thrombi in the HeartMate II Left Ventricular Assist Device. ASAIO J. 66, 992–999. https://doi.org/10.1097/MAT.0000000000001151

Ruggeri, Z.M., 2009. Platelet adhesion under flow. Microcirculation 16, 58–83. https://doi.org/10.1080/10739680802651477

Schwoch, G., Passow, H., 1973. Preparation and properties of human erythrocyte ghosts. Mol. Cell.

Biochem. 2, 197–218. https://doi.org/10.1007/BF01795474

Selmi, M., Chiu, W.C., Chivukula, V.K., Melisurgo, G., Beckman, J.A., Mahr, C., Aliseda, A., Votta, E., Redaelli, A., Slepian, M.J., Bluestein, D., Pappalardo, F., Consolo, F., 2019. Blood damage in Left Ventricular Assist Devices: Pump thrombosis or system thrombosis? Int. J. Artif. Organs 42, 113–124. https://doi.org/10.1177/0391398818806162

Sin, D., Kei, H., Miao, X., 2009. Surface coatings for ventricular assist devices. Expert Rev. Med. Devices 6, 51–60. https://doi.org/10.1586/17434440.6.1.51

Soares, J.S., Sheriff, J., Bluestein, D., 2013. A novel mathematical model of activation and sensitization of platelets subjected to dynamic stress histories. Biomech. Model. Mechanobiol. 12, 1127–1141. https://doi.org/10.1007/s10237-013-0469-0

Song, X., Throckmorton, A.L., Wood, H.G., Antaki, J.F., Olsen, D.B., 2004. Quantitative evaluation of blood damage in a centrifugal VAD by computational fluid dynamics. J. Fluids Eng. Trans. ASME 126, 410–418. https://doi.org/10.1115/1.1758259

Sorensen, E.N., Burgreen, G.W., Wagner, W.R., Antaki, J.F., 1999. Computational simulation of platelet deposition and activation: I. Model development and properties. Ann Biomed Eng 27, 436–448.

Spann, A.P., Campbell, J.E., Fitzgibbon, S.R., Rodriguez, A., Cap, A.P., Blackbourne, L.H., Shaqfeh, E.S.G., 2016. The Effect of Hematocrit on Platelet Adhesion : Experiments and Simulations. Biophysj 111, 577–588. https://doi.org/10.1016/j.bpj.2016.06.024

Strickland, K.C., Watkins, J.C., Couper, G.S., Givertz, M.M., Padera, R.F., 2016. Thrombus around the redesigned HeartWare HVAD inflow cannula: A pathologic case series. J. Hear. Lung Transplant. 35, 926–930. https://doi.org/10.1016/j.healun.2016.01.1230

Teuteberg, J.J., Cleveland, J.C., Cowger, J., Higgins, R.S., Goldstein, D.J., Keebler, M., Kirklin, J.K., Myers, S.L., Salerno, C.T., Stehlik, J., Fernandez, F., Badhwar, V., Pagani, F.D., Atluri, P., 2020. The Society of Thoracic Surgeons Intermacs 2019 Annual Report: The Changing Landscape of Devices and Indications. Ann. Thorac. Surg. 109, 649–660. https://doi.org/10.1016/j.athoracsur.2019.12.005

Tsubota, H., Ribeiro, R.V.P., Billia, F., Cusimano, R.J., Yau, T.M., Badiwala, M. V., Stansfield, W.E., Rao, V., 2017. Left ventricular assist device exchange: The Toronto General Hospital experience. Can. J. Surg. 60, 253–259. https://doi.org/10.1503/cjs.011316

Wang, L., Chen, Z., Zhang, J., Zhang, X., Wu, Z.J., 2020. Modeling Clot Formation of Shear-Injured Platelets in Flow by a Dissipative Particle Dynamics Method. Bull. Math. Biol. 82, 83. https://doi.org/10.1007/s11538-020-00760-9

Wu, W.-T., Jamiolkowski, M.A., Wagner, W.R., Aubry, N., Massoudi, M., Antaki, J.F., 2017. Multi-Constituent Simulation of Thrombus Deposition. Sci. Rep. 7, 42720. https://doi.org/10.1038/srep42720

Wu, W.-T., Zhussupbekov, M., Aubry, N., Antaki, J.F., Massoudi, M., 2020. Simulation of thrombosis in a stenotic microchannel: The effects of vWF-enhanced shear activation of platelets. Int. J. Eng. Sci. 147. https://doi.org/10.1016/j.ijengsci.2019.103206

Wu, W., Aubry, N., Massoudi, M., 2014a. On the coefficients of the interaction forces in a two-phase flow of a fluid infused with particles. Int. J. Non. Linear. Mech. 59, 76–82. https://doi.org/10.1016/j.ijnonlinmec.2013.11.006

Wu, W., Aubry, N., Massoudi, M., Kim, J., Antaki, J.F., 2014b. A numerical study of blood flow using mixture theory. Int. J. Eng. Sci. 76, 56–72. https://doi.org/10.1016/j.ijengsci.2013.12.001

Xu, S., Xu, Z., Kim, O. V, Litvinov, R.I., Weisel, J.W., Alber, M., 2017. Model predictions of deformation, embolization and permeability of partially obstructive blood clots under variable shear flow. J. R. Soc. Interface 14, 20170441. https://doi.org/10.1098/rsif.2017.0441

Yamane, T., 2016. How Do We Select Materials? BT - Mechanism of Artificial Heart, in: Yamane, T. (Ed.), . Springer Japan, Tokyo, pp. 51–55. https://doi.org/10.1007/978-4-431-55831-6_7

Yazdani, A., Li, H., Humphrey, J.D., Karniadakis, G.E., 2017. A General Shear-Dependent Model for Thrombus Formation. PLoS Comput. Biol. 13, 1–24. https://doi.org/10.1371/journal.pcbi.1005291

Ye, S., Johnson, C.A., Woolley, J.R., Murata, H., Gamble, L.J., Ishihara, K., Wagner, W.R., 2010. Simple surface modification of a titanium alloy with silanated zwitterionic phosphorylcholine or sulfobetaine modifiers to reduce thrombogenicity. Colloids Surfaces B Biointerfaces 79, 357–364. https://doi.org/10.1016/j.colsurfb.2010.04.018

Zhang, P., Gao, C., Zhang, N., Slepian, M.J., Deng, Y., Bluestein, D., 2014. Multiscale Particle-Based Modeling of Flowing Platelets in Blood Plasma Using Dissipative Particle Dynamics and Coarse Grained Molecular Dynamics. Cell. Mol. Bioeng. 7, 552–574. https://doi.org/10.1007/s12195-014-0356-5

Zhang, Y., Zhan, Z., Gui, X.M., Sun, H.S., Zhang, H., Zheng, Z., Zhou, J.Y., Zhu, X.D., Li, G.R., Hu, S.S., Jin, D.H., 2008. Design optimization of an axial blood pump with computational fluid dynamics. ASAIO J. 54, 150–155. https://doi.org/10.1097/MAT.0b013e318164137f

Zhao, R., Kameneva, M. V, Antaki, J.F., 2007. Investigation of platelet margination phenomena at elevated shear stress. Biorheology 44, 161–77.

Zhao, R., Marhefka, J.N., Shu, F., Hund, S.J., Kameneva, M. V, Antaki, J.F., 2008. Micro-Flow Visualization of Red Blood Cell-Enhanced Platelet Concentration at Sudden Expansion. Ann. Biomed. Eng. 36, 1130–1141. https://doi.org/10.1007/s10439-008-9494-z

Zheng, X., Yazdani, A., Li, H., Humphrey, J.D., Karniadakis, G.E., 2020. A three-dimensional phase-field model for multiscale modeling of thrombus biomechanics in blood vessels. PLOS Comput. Biol. 16, e1007709. https://doi.org/10.1371/journal.pcbi.1007709